\documentclass[12pt]{article}
\usepackage{graphicx} 
\begin{document}
\centerline{\bf Analysing and controlling the tax evasion dynamics via majority-vote model }

\bigskip
\centerline{F.W.S. Lima }
 
\bigskip
\noindent

Departamento de F\'{\i}sica, 
Universidade Federal do Piau\'{\i}, 64049-550, Teresina - PI, Brazil.\\

\medskip
  e-mail: fwslima@gmail.com, wel@ufpi.edu.br
\bigskip
 
{\small Abstract: Within the context of agent-based Monte-Carlo simulations, we study the well-known majority-vote model (MVM) with noise applied to tax evasion on simple square lattices, Voronoi-Delaunay random lattices, Barabasi-Albert networks, and Erd\"os-R\'enyi random graphs. In the order to analyse and to control the fluctuations for tax evasion in the economics model proposed by Zaklan, MVM is applied  in the neighborhod of the noise critical $q_{c}$. The Zaklan model  had been studied recently using the equilibrium Ising model. Here we show that the Zaklan model is robust and can be reproduced also through the nonequilibrium MVM on various topologies.}
 
 
\bigskip

{\bf 1. Introduction}

The Ising model \cite{a3,onsager} has been used during long time as a "toy model" for diverses objectives, as to test and to improve new algorithms and methods of high precision for calculation of critical exponents in Equilibrium Statistical Mechanics  using the Monte Carlo method
as Metropolis \cite{me}, Swendsen-Wang \cite{s-w}, Wang-Landau \cite{w-l} algorithms, Single histogram  \cite{f-s} and Broad histogram \cite{pmco} methods. The Ising model was already applied decades ago to explain how a school of fish aligns into one direction for swimming \cite{callen} or how workers decide whether or not to go on strike \cite{galam}. In the Latan\'e model of Social Impact \cite{latané} the Ising model has been used to give a consensus, a fragmentation into many different opinions, or a leadership effect when a few people change the opinion of lots of others. To some extent the voter model of Liggett \cite{ligg} is an Ising-type model: opinions follow the majority of the neighbourhood, similar to Schelling \cite{Schel}, all these cited model and others can be found out in \cite{book}. 

Realistic models of tax evasion appear to be necessary because tax evasion remain to be a major predicament facing governments \cite{K,JA,L,JS}. Experimental evidence provided by G\"achter \cite{Ga} indeed suggests that tax payers tend to condition their decision regarding whether to pay taxes or not on the tax evasion decision of the members of their group. Frey and Torgler \cite{FT} also provide empirical evidence on the relevance of conditional cooperation for tax morale. Following the same context, recently, Zaklan et al. \cite{zaklan,zaklan1} developed an economics model to study the problem of tax evasion dynamics using the Ising model through Monte-Carlo simulations with the Glauber and heatbath algorithms ( that obey detailed balance  - equilibrium) to study the proposed model.

G. Grinstein et al. \cite{g} have argued that nonequilibrium stochastic spin systems on regular square lattices with up-down symmetry fall into the universality
class of the equilibrium Ising model \cite{g}. This conjecture was confirmed for various Archimedean lattices and 
in several models that do not obey detailed balance \cite{C,J,M,mario,lima01}. The majority-vote model (MVM) is a nonequilibrium model proposed by M.J. Oliveira in $1992$ and defined by stochastic dynamics with local rules and with up-down symmetry on a regular lattice shows a second-order phase transition with critical exponents $\beta$, $\gamma$, $\nu$ which characterize  the system in the vicinity of the phase transition identical \cite{mario,a1} with those of the equilibrim Ising model \cite{a3} for regular lattices. Lima et al. \cite{lima0} 
studied MVM  on Voronoi-Delaunay random lattices
with periodic boundary conditions. These lattices posses natural quenched
disorder in their connections. They showed that presence of quenched 
connectivity disorder is enough to alter the exponents $\beta/\nu$
and $\gamma/\nu$ from the pure model and therefore that is a relevant term to
such non-equilibrium phase-transition with disagree with the arguments of G. Grinstein et al. \cite{g}. Recently, simulations on both {\it undirected} and {\it directed} scale-free 
networks \cite{newman,sanchez,ba1,alex,sumour,sumourss,lima}, random graphs \cite{erdo,er2} and social networks \cite{er,er1,DS}, have attracted interest of researchers from various areas. These complex networks have been 
studied extensively by Lima et al. in the context of magnetism (MVM, Ising, and Potts model) \cite{lima1,lima2,lima3,lima4,lima5,lima6}, econophysics models \cite{zaklan1,lima8} and sociophysics model \cite{lima9}. 

In the present work, we study the behavior of the tax evasion on two-dimensional square lattices using MVM dynamics and furthermore add a policy makers's tax enforcement mechanism consisting of two components: a probability of an audit each person is subject to in every period and a length of time detected tax evaders remain honest. We aim to extend the study of Zaklan et al. \cite{zaklan,zaklan1}, which illustrates how different levels of enforcement affect the tax evasion over time, as an alternative model of  nonequilibrium to the Ising model that is capable of reproduce the same results for analysis and control the fluctuation of the tax evasion. This MVM shows that the Zaklan model is very robust for equilibruim and nonequilibrium models and also for various topologies. This does not always happen with other models. 

The remainder of our paper is organised as follows. In section 2, we present the Zaklan model evolving with dynamics of MVM. In section 3 we make an analysis of tax evasion dynamics with the Zaklan model on two-dimensional square lattices using 
MVM for their temporal evolution under different enforcement regimes; we discuss the results obtained. In section 4 we show that the MVM model also is capable to control the different levels of the tax evasion analysed in section 3, as it was made by Zaklan et al. \cite{zaklan1} using Ising models. We use the enforcement mechanism cited above on various structures: square lattice, Voronoi-Delaunay random lattice, Barab\'asi-Albert network and Erd\"os-R\'enyi graph; we discuss the resulting tax evasion dynamics. Finally in section 5 we present our conclusions about the study of the Zaklan model using MVM.
 
\bigskip

{\bf 2. Zaklan model }

On a square lattice where each site of the lattice is inhabited, at a time step, by an  agent with "voters" or spin variables ${\sigma}$ taking the values $+1$ representing an honest tax payer, or $-1$ trying to at least partially escape her tax duty. Here is assumed that initially everybody is honest. Each period individuals can rethink their behavior and have the opportunity to become the opposite type of agent they were in previous period. In each time period the system evolves by a single spin-flip dynamics with a probability $w_{i}$ given by
\begin{equation}
%\begin{center}
w_{i}(\sigma)=\frac{1}{2}\biggl[ 1-(1-2q)\sigma_{i}S\biggl(\sum_{\delta
=1}^{k_{i}}\sigma_{i+\delta}\biggl)\biggl],
%\end{center}
\end{equation}
where $S(x)$ is the sign $\pm 1$ of $x$ if $x\neq0$, $S(x)=0$ if $x=0$, and the 
summation runs over all $k_i$ nearest-neighbour sites $\sigma_{i+\delta}$ of $\sigma_{i}$. In this model an agent assumes the value $\pm 1$ depending on the opinion of the majority of its neighbors. The control noise parameter $q$ plays the role of the temperature in equilibrium systems and measures the probability of aligning antiparallel to the majority of neighbors. Then various degrees of homogeneity regarding either position are possible. An extremely homogenous group is entirely made of honest people or tax evaders, depending of the sign $S(x)$ of the majority of neighbhors. If $S(x)$ of the neighbors is zero the agent $\sigma_{i}$ will be honest or evader in the next time period with probability $1/2$. We further introduce a probability of an efficient audit ($p$). Therefore, if tax evasion is detected, the agent must remain honest for a number $k$ of time steps. Here, one time step is one sweep through the entire lattice. 
\bigskip

{\bf 3. Analysing tax evasion dynamics }

Here then as  Zaklan et al \cite{zaklan} we also use square lattices with lattice size $L=1000\; (L \times L=10^{6}$ sites) and simulate tax evasion dynamics for various values of $q$ and various punishments ($k$). When $k$ is zero, no punishment is present and the model describes the baseline majority-vote model. Several degrees of punishment are introduced for different noise $(q)$ or "Social Temperature", by setting $k$ consecutively equal to $10$ and $50$ periods for all considered levels of the noise. The probability of an audit is sequentially increased, in steps, from $0$ to $100$ percent. For a given probability of an audit the dynamics of the tax evasion (measured as portion of the entire population) is shown over $300$ time steps. Here, we  have observed the behavior of tax evasion dynamics on a square lattice for high $q=0.70$ and low $q=0.050$ noise values of the critical value for square lattices($q_{c}=0.075$) the MVM model. For both  periods time punishment, $K=50$ and $10$,  we also use two probability of an audit is either at a realistic level ($p=0.05$) or at a rather high level ($p=0.9$) values in this model. The tax evaders have the greatest influence to turn honest citizens into tax evaders if they contitute a majority in given neighbourhood. If the majority evades, one is likely to also evade. On other way, if the majority of agents in the vicinity are honest, the respective individual is likely to become a honest citizen even though it has been a tax evader before. Therefore, in the fig. 1-3, have been showed that on square lattice the MVM model (nonequilibrium model) is capable to reproduce the same behavior and results for analysis of tax evasion levels, as Zaklan et al. \cite{zaklan} made using the Ising model (equilibrium model) on square lattices. How strong the influence from the neighbourhood is can be controlled by adjusting or setting the noise parameter, $q$, next to ($q_{c}$).

\begin{center}
--- Figure 1 goes about here ---
\end{center}                
\noindent
\begin{figure}[hbt]
%\begin{center}
\includegraphics[angle=0,scale=0.65]{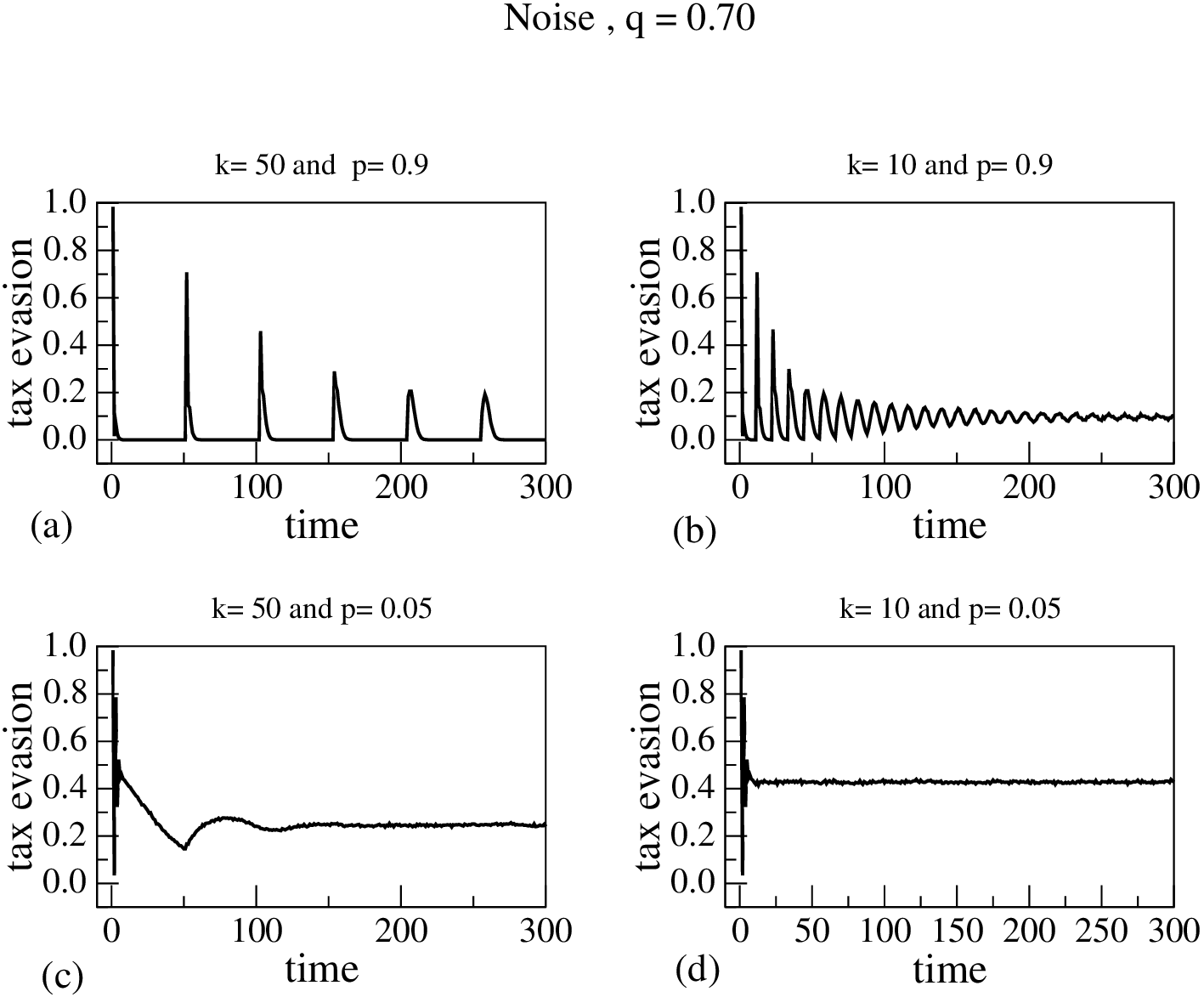}
%\end{center}
\caption{Tax evasion dynamics for  high "social temprature" $q=0.70$ ($q_{c}=0.075$) and $300$ time step, $L=1000\; (10^6$ sites), with length of punishment at either $k=50$ or $10$ periods and two different probabilites of an audit, i.e. $p=0.05$ and $p=0.9$ .}
\end{figure} 
In figure 1 we plot tax evasion for high noise $q=0.70$, for punishment periods $k=50$ and 10. If the penalty is high enough, $k=50$, tax evasion can be reduced to $0\%$ in the short run, given that the probability of an audit $p$ is sufficiently high. In the case of $k=50$ and $p=90\%$, picture (a) of Fig. 1, within only a few time steps each individual eventually is forced to remain honest. This happens, because the spins flip more often at this noise, which is too high compared to the critical noise of this model in the square lattice ($q_{c}=0.075$) \cite{mario}, where $q_{c}$ is the critical noise for phase transition of MVM on square lattice. In this picture (a) we observe peaks of different levels of the tax evasion, that result from the fact that $90\%$ of the initial large number of tax evaders gets caught and after $k$ iterations becomes free to decide whether to evade or not. Moreover consecutive peaks in non-compliance diminish less over time and it takes longer until perfect non-compliance is established, further out on the time  scale at which the evolution of tax evasion is considered. Therefore, in our simulation using MVM the perfect compliance is not attained in agreement with the results observed for Zaklan et al. \cite{zaklan} using the Ising model where an equilibrium level of about $2\%$ non-compliance is attained.

If punishment is set equal to $k=10$ periods (picture (b) of Fig. 1) tax evasion reaches $0\%$ in the first ten time steps and finally comes to rest at a level of $10\%$. When the probability of an audit $p=0.05$ is much lower the tax evasion level reaches values for non-compliance of about $24\%$ and $43\%$ in pictures (c) and (d), respectively, of Fig. 1. These two pictures show well that punishment is a suitable enforcement mechanism when the probability of an audit is set to a realistic level.
\begin{center}
--- Figure 2  and 3 goes about here ---
\end{center}

\begin{figure}[hbt]
%\begin{center}
\includegraphics[angle=0,scale=0.65]{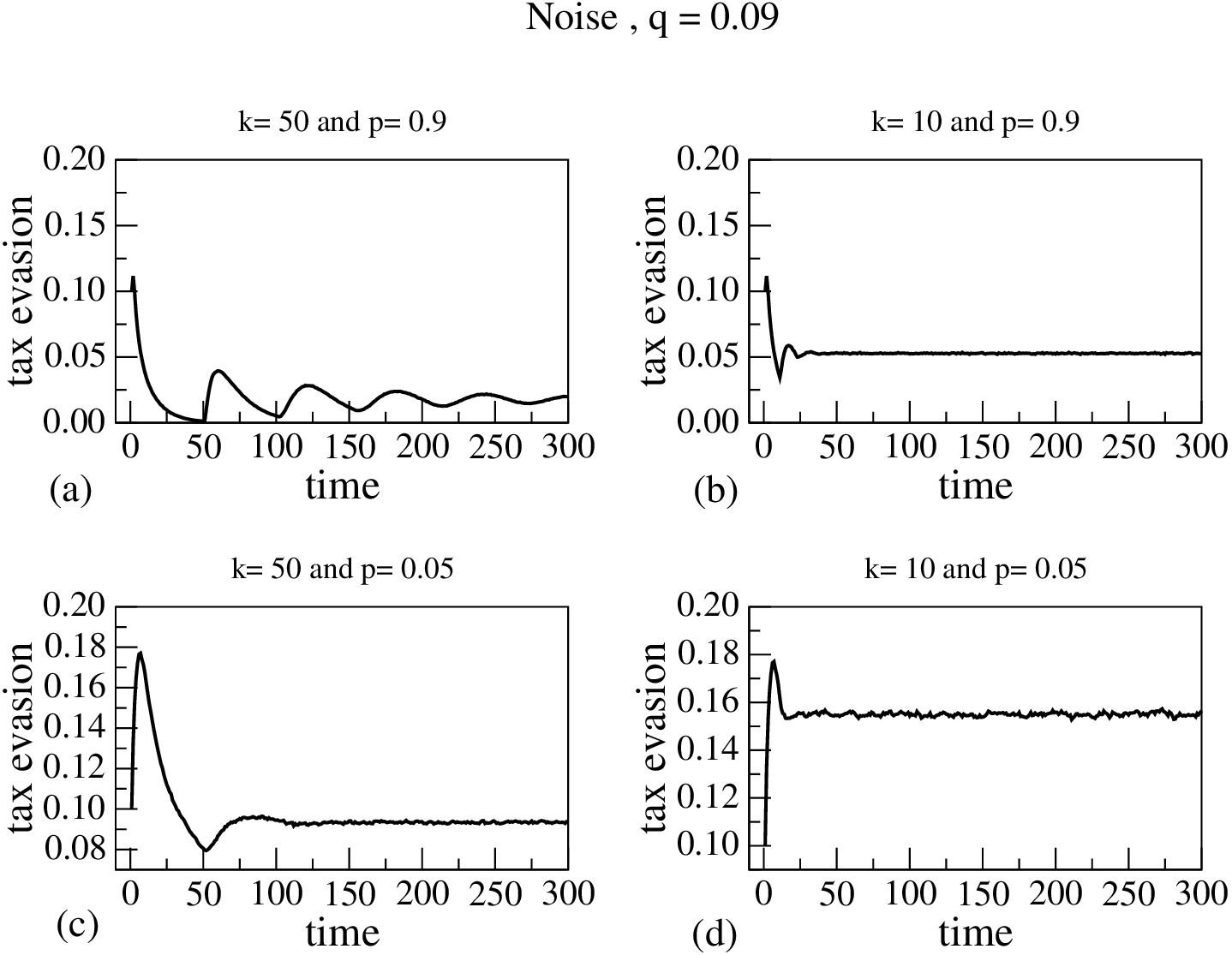}
%\end{center}
\caption{Tax evasion dynamics for  "social temperature" $q=0.090$ ($q_{c}=0.075$). The same simulation design as in figure 1.}
\end{figure} 
\begin{figure}[hbt]
%\begin{center}
\includegraphics[angle=0,scale=0.65]{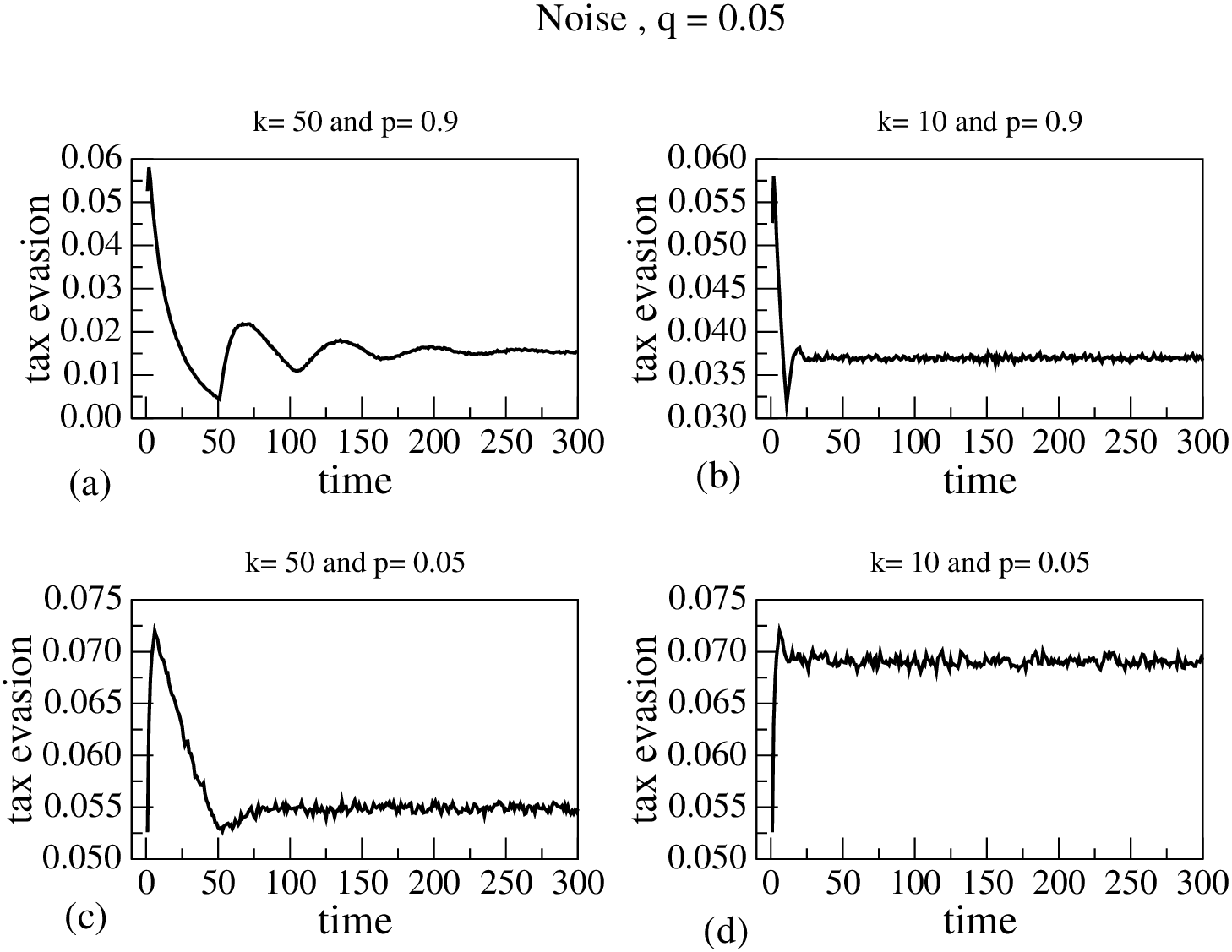}
%\end{center}
\caption{Tax evasion dynamics for  "social temprature" $q=0.050$. The same simulation design as in figure 1.}
\end{figure} 
We also consider two other noises above $q_{c}$, $q=0.090$, and below it, $q=0.050$. At $q=0.090$ the individual becomes tax evader more slowly, because the tax evasion problem is less pronounced already from the beginning compared to higher noises, because spins flip less frequently at lower noises, the same enforcement mechanisms may work less efficintly in the short run than at higher noise. Therefore as shown in Fig. 2 at $q=0.090$ with  $k=50$ and $p=0.9$ one clearly sees that evasion cannot be reduced to zero percent any more. At $q=0.050$ the shape of tax evasion is similar to $q=0.090$, but the level of tax evasion is smaller, because spins flip less frequently at lower noises $q=0.050$ as shown in Fig. 3, again enforcement mechanisms cannot work efficiently in the short run as opposed to higher noise ($q=0.70$). At such low noise individuals seldomly decide to become non-compliant because their vicinity which is mostly compliant on average exerts strong influence on them to be honest as well. Therefore the equilibrium level of tax evasion is smaller than at higher noises.

\bigskip

{\bf 4. Controlling the tax evasion dynamics }

Here, we first will present the  baseline case $k=0$, i.e., no use of enforcement, for different network structure. We use for square lattice $q=0.95 q_{c}$, for Voronoi-Delaunay lattice at $q=q_{c}=0.117$ (critical noise for MVM model on Voronoi-Delaunay lattice), for Barab\'asi-Albert 
$q=0.9 q_{c}=0.306$ and Erd\"os-R\'enyi $q=0.9 q_{c}=0.181$. All simulation are performed over $20,000$ time steps, it is showed in the fig. 4.

For very low noises the part of autonomous decisions almost completely disappears. The individuals then base their decision solely on what most of their neighbours do. A rising noise has the opposite effect. Individuals then decide more autonomously. For the MVM model it is known that for $q>q_{c}$, half of the people are honest and other half cheat, while for $q<q_{c}$ states correlated on cheating or compliance prevail for most of the time. Because this behavior we set some values closed to $q_{c}$, where the case that agents distribute in equal proportions onto the two alternatives is excluded. Then have been setting the noise parameter, $q$, closed to ($q_{c}=0.075$) on square lattice, as suggested in the section 3, and ranging the degrees of punishment ($k=1$, $10$ and $50$) and audit probability rate ($p=0.5\%$, $10\%$ and $90\%$). Therefore, if tax evasion is detected, the  enforcement mechanism ($p$) and the period time of punishment $k$ are triggered in order of to control the tax evasion level. The individual remain honests for a certain number of periods, as explained before in section 2 and 3. We also 	
extend our study to other networks as the Voronoi-Delaunay lattices, Barab\'asi-Albert networks and Erd\"os-R\'enyi graphs with $N=400$ sites. As before the initial configurations is with all honest agents ($\sigma_{i}=+1$) at fixed "Social Temperature" ($q$). Here, have been performed simulations of $10,000$ to $20,000$ time steps.
\begin{center}
--- Figure 4 goes about here ---
\end{center}

\begin{figure}[hbt]
%\begin{center}
\includegraphics[angle=0,scale=0.65]{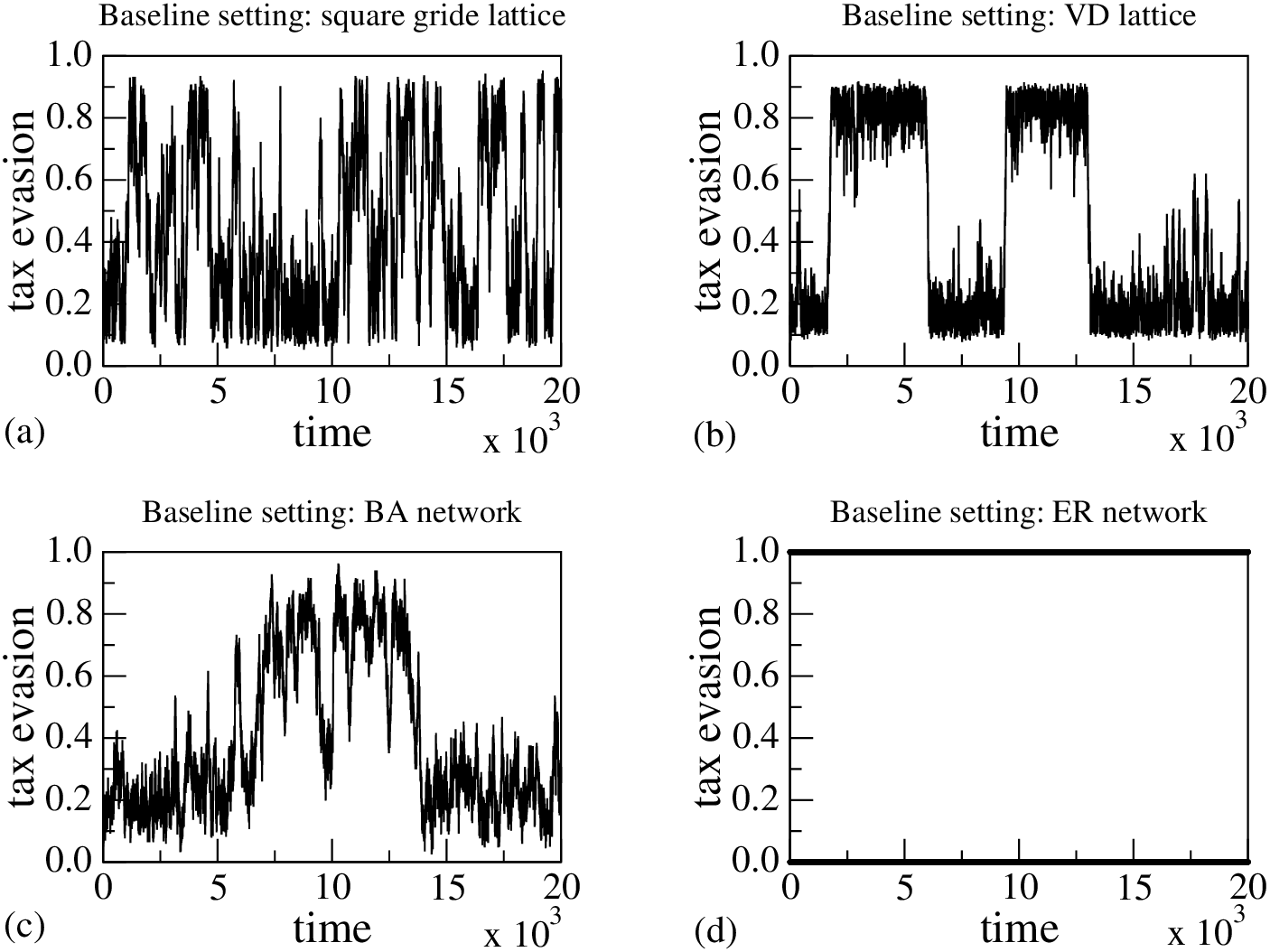}
%\end{center}
\caption{Baseline case for different network structure. Where we use for square lattice $q=0.95 q_{c}$, for Voronoi-Delaunay lattice at $q=q_{c}=0.117$ (critical noise for MVM model on Voronoi-Delaunay lattice), for Barab\'asi-Albert 
$q=0.9 q_{c}=0.306$ and Erd\"os-R\'enyi $q=0.9 q_{c}=0.181$. All simulation are performed over $20,000$ time steps.}
\end{figure} 

In Fig. 4 we plot the baseline case $k=0$, i.e., no use of enforcement, for the square lattice (a), Voronoi-Delaunay (b), Barab\'asi-Albert (c), and Erd\"os-R\'enyi (d) for dynamics of the tax evasion over $20,000$ time steps. Although everybody is honest initially, it is impossible to predict which level of tax compliance will be reached at some time step in the future.
\begin{center}
--- Figure 5 goes about here ---
\end{center}

\begin{figure}[hbt]
%\begin{center}
\includegraphics[angle=0,scale=0.65]{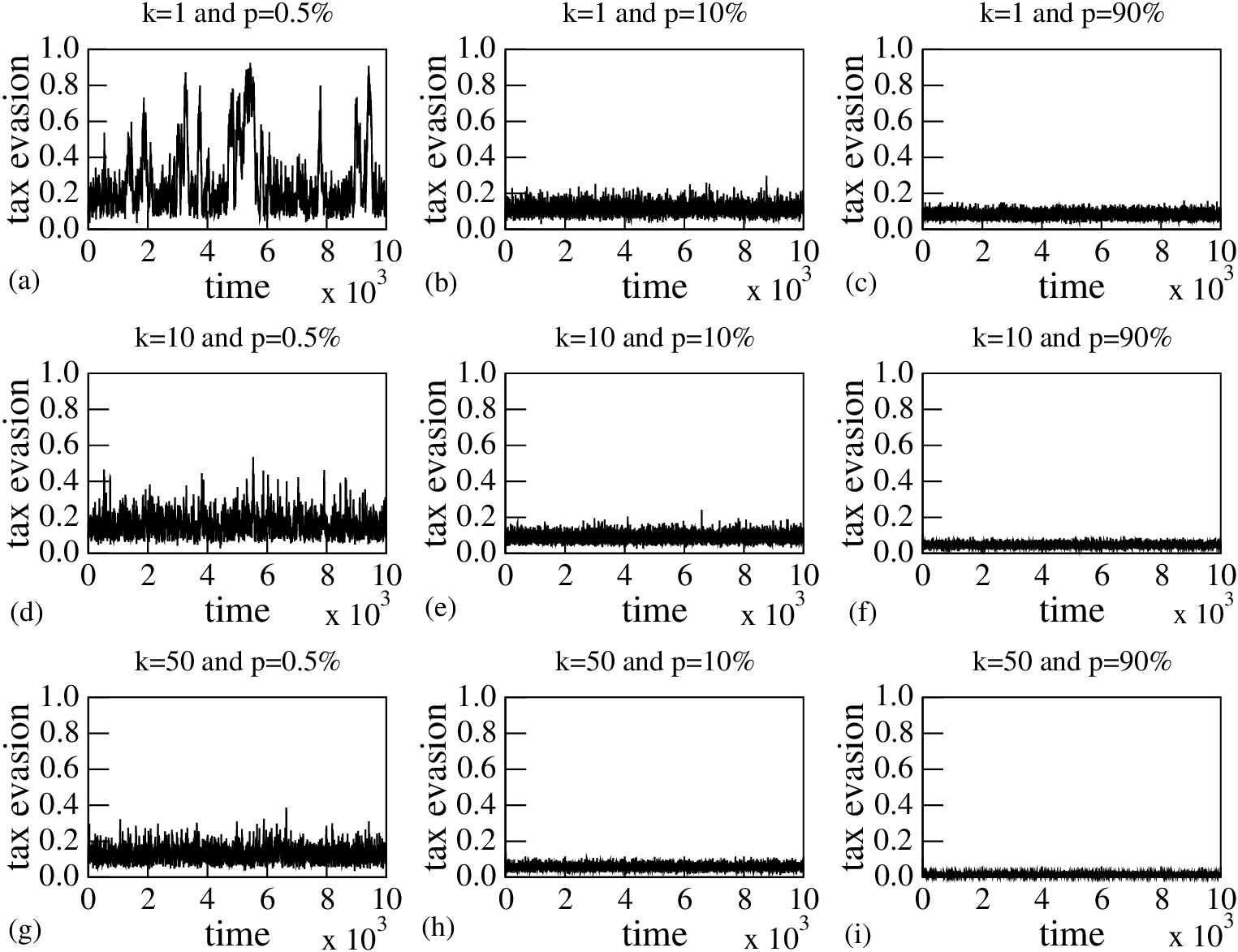}
%\end{center}
\caption{The square lattice model of tax evasion with various degrees $p$ 
of enforcement. $q=0.95 q_{c}$ and $20,000$ time steps.}
\end{figure} 
Figure 5 illustrates different simulation settings for square lattice, for each considered combination of degree of punishment ($k=1$, $10$ and $50$) and audit probability rate ($p=0.5\%$, $10\%$ and $90\%$), where the tax evasion is plotted over 10,000 time steps. Here we show that even a very small level the enforcement ($p=0.5\%$ and $k=1$) suffices to reduce
fluctuations in tax evasion and to establish mainly compliance. Both a rise in audit probability (greater $p$) and higher penalty (greater $k$) work to flatten the time series of tax evasion and to shift the band of possible non-compliance values towards more compliance. However, the simulations show that even extreme enforcement measures ( $p=90\%$ and $k=50$) cannot fully solve the problem of tax evasion.
\begin{center}
--- Figure 6 goes about here ---
\end{center}

\begin{figure}[hbt]
%\begin{center}
\includegraphics[angle=0,scale=0.65]{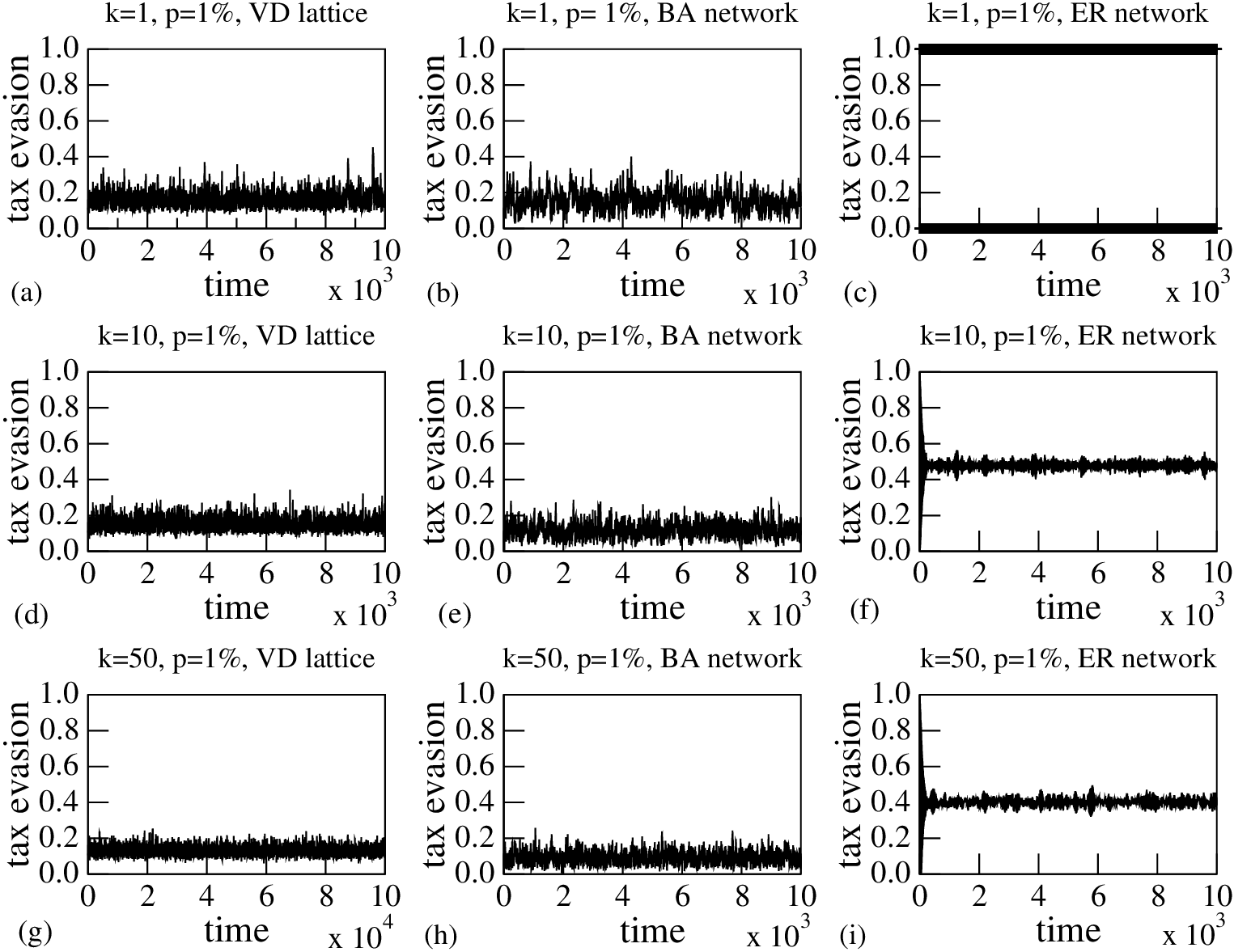}
%\end{center}
\caption{The first column displays tax evasion for different enforcement regimes for Voronoi-Delaunay network ((a), (d), and (g)). The next two column depict the tax evasion for Barab\'asi-Albert network ((b), (e), and (h)) and Erd\"os-R\'enyi graph((c), (f), and (i)). Again, we use $10,000$ time steps.}
\end{figure} 
In Fig. 6 we display tax evasion for Voronoi-Delaunay random lattice, Barab\'asi-Albert network, and Erd\"os-R\'enyi random graph for different enforcement for $k=1$, $10$, and $50$ with the same audit probability $p=1\%$. We observe for the Voronoi-Delaunay random lattice and Barab\'asi-Albert network that the tax evasion level decreases with increasing time periods $k$ of punishment. For the case of the  Erd\"os-R\'enyi random graph the tax evasion fluctuations for $k=1$ and $p=1\%$ , case (c), is identical to the baseline case for this random graph, only for $k=10$ and $50$ at $p=1 \%$ the tax evasion level decreases to about $48\%$ (f) and $40\%$ (i), showing that the control for tax evasion for small audit probability $p=1\%$ is more difficult to reach in these Erd\"os-R\'enyi random graphs.

\begin{center}
--- Figure 7 goes about here ---
\end{center}

\begin{figure}[hbt]
%\begin{center}
\includegraphics[angle=0,scale=0.65]{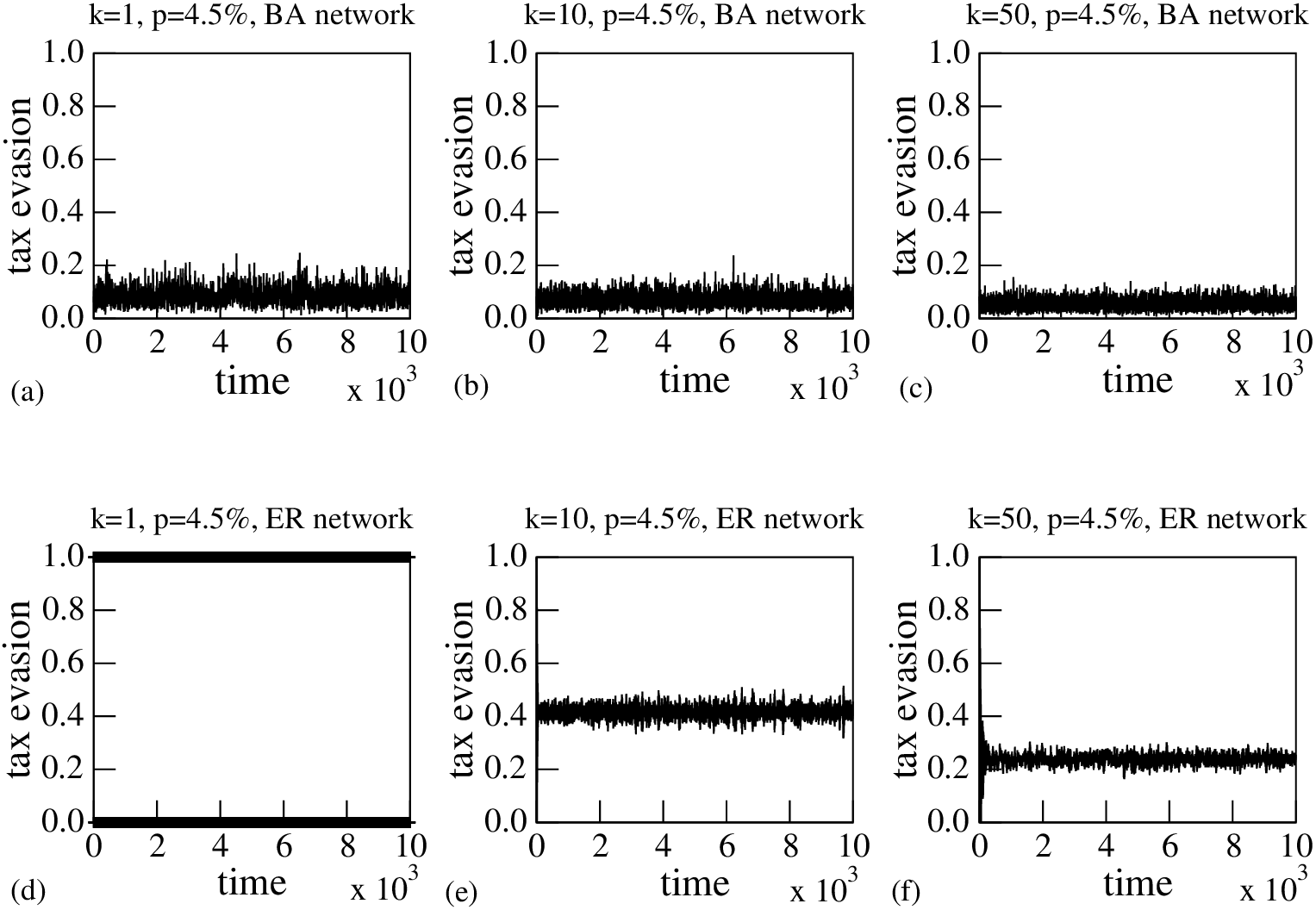}
%\end{center}
\caption{The first line display the resulting tax evasion for different enforcement regimes for Barab\'asi-Albert networks ((a), (b), and (c)) for degrees of punishment $k=1$, $10$ and $50$ and audit probability $p=4.5\%$,. The second line depicts tax evasion for Erd\"os-R\'enyi graphs( (d), (e), and (f)) for the same values the case as line 1. Again, we use $10,000$ time steps.}
\end{figure} 

In Fig. 7 we plot tax evasion for Barab\'asi-Albert networks and Erd\"os-R\'enyi random graphs, again for different enforcement $k=1$, $10$, and $50$, but now with audit probability $p=4.5\%$. For Barab\'asi-Albert networks the tax evasion level decreases with increasing audit probability $p$ showing that an increase of the audit probability favors the control of tax evasion. Again, for $k=1$ and $p=4.5\%$ (d) the tax evasion for the Erd\"os-R\'enyi graphs is identical for case baseline and the case (c) of Fig. 6, showing that for these random graphs the time period $k$ of punishment is important to control tax evasion.

\bigskip
{\bf 5. Conclusion}

In summary, tax evasion can vary widely across nations, reaching extremely high values in some developing countries. Wintrobe and  G\"erxhani \cite{WGa} explains the observed higher level of tax evasion in generally less developed countries with a lower amount of trust that people have in governmental institutions. In this work we show that Zaklan's model is very robust for analysis and control of tax evasion, using dynamics with equilibrium \cite{zaklan1}, Ising model, and also nonequilibrium, MVM. The Zaklan model is also found to be robust in various topologies used here.

The author thank Dietrich Stauffer for the revision of this paper. We also acknowledge the
Brazilian agency FAPEPI (Teresina-Piau\'{\i}-Brasil) for  its financial support. This
work also was supported the system SGI Altix 1350 the computational park
CENAPAD.UNICAMP-USP, SP-BRAZIL.

\end{document}